\documentclass[12pt]{article}
\usepackage{epsfig}

\usepackage{graphicx}
\newcommand{\beq}{\begin{equation}}
\newcommand{\eeq}{\end{equation}}
\newcommand{\beqn}{\begin{eqnarray}}
\newcommand{\eeqn}{\end{eqnarray}}
\newcommand{\beqs}{\begin{eqnarray*}}
\newcommand{\eeqs}{\end{eqnarray*}}
\begin{document}

\title{\Large \bf Analytic properties of
high energy \\production amplitudes
in $N=4$ SUSY}
\author{\large L.~N. Lipatov$^{1,2}$
\bigskip \\
{\it
$^1$~II. Institut f\"{u}r Theoretische Physik, Hamburg University, Germany} \\
{\it  $^2$~St. Petersburg Nuclear Physics Institute, Gatchina, Russia}}

\maketitle

\vspace{-10cm}
\begin{flushright}
$~~$\\
DESY-10-125
\end{flushright}
\vspace{8cm}

\abstract{We investigate analytic properties of the six point planar amplitude
in $N=4$ SUSY at the multi-Regge kinematics for final state particles. For inelastic processes the
Steinmann relations play an important  role because they give a possibility to fix the phase
structure of the Regge pole and Mandelstam cut
contributions. These contributions have  the M\"{o}bius invariant form in the transverse
momentum subspace. The analyticity and 
factorization constraints
allow us to reproduce the two-loop correction
to the 6-point BDS amplitude
in $N=4$ SUSY obtained earlier in the leading logarithmic approximation
with the use of the $s$-channel unitarity. The exponentiation hypothesis
for the remainder function in the multi-Regge kinematics
is also investigated. The 6-point amplitude in LLA can be completely reproduced
from the BDS ansatz with the use of the analyticity and Regge factorization.}

\section{Introduction}

The elastic scattering amplitude in QCD at high energies $\sqrt{s}$ and fixed
momentum transfers $q=\sqrt{-t}$ for the transition $AB\rightarrow A'B'$
with the definite particle helicities
$\lambda _i$ in the leading
logarithmic approximation (LLA) has the Regge form~\cite{BFKL}
\begin{equation}
A_{2\rightarrow 2}=2\,g\delta _{\lambda _{A}\lambda _{A'}}
T_{AA'}^c\frac{s^{1+\omega (t)}}{t}\,g\,T_{BB'}^c
\,\delta _{\lambda _{B}\lambda _{B'}}\,,\,\,t=-\vec{q}^{{2}}.
\end{equation}
The gluon Regge trajectory $j(t)=1+\omega (t)$ in LLA is given below
\begin{equation}
\omega (-\vec{q}^{2})=-\frac{\alpha_{s} N_c}{(2\pi )^2}\,
(2\pi \mu )^{2\epsilon}\,\int
d^{2-2\epsilon }k
\,\frac{
\vec{q}^{2}}{\vec{k}^{2}(\vec{q}-{k})^{2}}\approx
-\,a\,\left(\ln
\frac{\vec{q}^{2}}{\mu ^2}-\frac{1}{\epsilon}\right)\,,
\end{equation}
where we introduced the dimensional regularization ($D=4-2\,\epsilon$)
and the renormalization point $\mu$ for the t' Hooft coupling
constant
\begin{equation}
a=\frac{\alpha _{s}\,N_c}{2\pi }\,\left(4\pi e^{-\gamma}\right)^\epsilon \,.
\end{equation}
The gluon trajectory is also known in the next-to-leading approximation at
QCD~\cite{trajQCD} and in SUSY gauge models~\cite{trajN4}.

For finding the total cross-section  in LLA it is enough to calculate the
production amplitudes in
the multi-Regge kinematics for the final state gluons. They have
the simple factorized form~\cite{BFKL}
\begin{eqnarray}
&&\hspace{-1cm}A_{2\rightarrow 2+n} ~=~ \nonumber\\
&&\hspace{-0.8cm}-2\,s\,
g \, \delta _{\lambda _A\lambda _{A'}}\,T^{c_1}_{AA'}
\frac{s_1^{\omega (-\vec{q}_1^2)}}{\vec{q}_1^2}gC_{\mu}(q_2,q_1)
e^*_\mu (k_1)T^{d_1}_{c_2c_1}\frac{s_2^{\omega (-\vec{q}_2^2)}}{\vec{q}_2^2}
...\frac{s_{n+1}^{\omega (-\vec{q}_{n+1}^2)}}{\vec{q}_{n+1}^2}
g \, \delta _{\lambda _B\lambda _{B'}}\,T^{c_{n+1}}_{BB'}\,,
\label{MRKcinematica}
\end{eqnarray}
where
\begin{equation}
s=(p_A+p_B)^2\gg s_r=(k_r+k_{r-1})^2\gg \vec{q}_r^2\,,\,\,\,k_r=q_{r+1}-q_r\,.
\end{equation}
The matrices $T^{a}_{b c}$ are the generators of the $SU(N_c)$ gauge
group in the adjoint representation and $C_\mu (q_r,q_{r-1})$ are the
effective Reggeon-Reggeon-gluon vertices. In the case when the polarization
vector $e_{\mu}(k_1)$ describes a produced gluon with a definite helicity
one can obtain~\cite{effmult}
\begin{equation}
C\equiv C_\mu
(q_2,q_1)\,e^*_{\mu}(k_1)=\sqrt{2}\,\frac{q_2^*q_1}{k^*_1}\,,
\label{helicityproduction}
\end{equation}
where the complex notation $q=q_x+iq_y$ for the two-dimensional transverse
vectors was used.

The elastic scattering amplitude with vacuum quantum numbers in the
$t$-channel can be calculated with the use of $s$-channel
unitarity~\cite{BFKL}. In this approach the Pomeron appears as a composite
state of two Reggeized gluons. It is convenient to present
transverse gluon coordinates in a complex form together with their canonically
conjugated momenta
\begin{equation}
\rho
_{k}=x_{k}+iy_{k}\,,\,\,\rho
_{k}^{\ast
}=x_{k}-iy_{k}\,,\,\,p_{k}=i
\frac{\partial }{\partial \rho
_{k}}\,,\,\,p_{k}^{\ast }=
i\frac{\partial }{\partial \rho
_{k}^{\ast }}\,.
\end{equation}
In the coordinate representation the Balitsky-Fadin-Kuraev-Lipatov (BFKL)
equation for the Pomeron wave function can be written as
follows~\cite{BFKL}
\begin{equation}
E\,\Psi
(\vec{\rho}_{1},\vec{\rho}_{2})=
H_{12}\,\Psi (\vec{\rho}_{1},\vec{%
\rho}_{2})\;,\,\,\Delta
=-\frac{\alpha _{s}N_{c}}{2\pi
}\,\min \,E\,,
\end{equation}
where $\Delta$ is the Pomeron intercept entering in the expression $\sigma_t\sim s^\Delta$ for
the high energy asymptotics of the total cross section. The BFKL Hamiltonian has a
simple operator representation~\cite{int1}
\begin{equation}
H_{12}=\ln
\,|p_{1}p_{2}|^{2}+\frac{1}{p_{1}p_{2}^{\ast
}}(\ln \,|\rho _{12}|^{2})\,p_{1}p_{2}^{\ast
}+\frac{1}{p_{1}^{\ast
}p_{2}}(\ln \,|\rho
_{12}|^{2})\,p_{1}^{\ast
}p_{2}-4\psi (1)
\end{equation}
with $\rho _{12}=\rho _1-\rho
_2$ and $\psi (x)=\Gamma '(x)/\Gamma (x)$. The kinetic energy is proportional to the sum
of two gluon Regge
trajectories $\omega (-|p|_i^2)$ ($i=1, 2$).
The potential energy $\sim \ln \,|\rho_{12}|^{2}$ is related to the product of
two gluon production vertices $C_\mu$. The Hamiltonian is invariant under the
M\"{o}bius transformation~\cite{moeb}
\begin{equation}
\rho _{k}\rightarrow
\frac{a\rho _{k}+b}{c\rho
_{k}+d}\,,
\end{equation}
where $a,b,c$ and $d$ are complex parameters. The eigenvalues of two
Casimir operators are expressed in terms of the
corresponding conformal weights
\begin{equation}
m=\frac{1}{2}+i\nu
+\frac{n}{2}\,,\,\,\widetilde{m}=\frac{1}{2}+i\nu
-\frac{n}{2}
\end{equation}
and for the principal series of unitary representations of $SL(2,C)$ the parameter $\nu$ is
real and $n$ is integer.

It turns out, that the BFKL pomeron has the positive intercept $\Delta=g^2N_c\ln 2\,/\pi^2$ in LLA,
which is not compatible with the $s$-channel unitarity. To restore the unitarity
one should take into account the diagrams with
an arbitrary number of Reggeized gluons in the $t$-channel. The composite states
of these gluons are described by the
Bartels-Kwiecinski-Praszalowicz
(BKP) equation~\cite{BKP}.
In the $N_c\rightarrow \infty$ limit the corresponding Hamiltonian has the property of
holomorphic separability~\cite{separ}
\begin{equation}
H=\frac{1}{2}\sum _kH_{k,k+1}=\frac{1}{2}
(h+h^*)\,,\,\,[h,h^*]=0\,.
\end{equation}
The holomorphic
Hamiltonian is a sum of the BFKL hamiltonians $h_{k,k+1}$
\begin{equation}
h=\sum _kh_{k,k+1}\,,\,\,h_{12}=\ln
(p_{1}p_{2})+\frac{1}{p_{1}}\,(\ln
\rho _{12})\,p_{1}+
\frac{1}{p_{2}}\,(\ln \rho
_{12})\,p_{2}-2\psi (1)\,.
\end{equation}
Consequently, the wave function $\Psi $ has properties of holomorphic
factorization~\cite{separ} and  duality
symmetry under the transformation~\cite{dual}
\begin{equation}
p_i\rightarrow \rho _{i,i+1}\rightarrow p_{i+1}\,.
\end{equation}
Moreover, in the holomorphic and anti-holomorphic
sectors, there are integrals of motion
commuting among themselves and with $h$~\cite{int1, int}.

The integrability of BFKL dynamics was firstly demonstrated
in~\cite{int}. It is
related to the fact that $h$ in LLA coincides with a local Hamiltonian of the
Heisenberg spin model \cite{LiFK}.

In the next-to-leading logarithmic approximation
the integral kernel for the BFKL equation was
constructed in Refs.~\cite{trajN4,FL}.
Due to its M\"{o}bius
invariance, solutions of the BFKL
and BKP equations can be classified by the anomalous
dimension $\gamma =
\frac{1}{2}+i\nu$ of twist-2 operators and
the conformal spin $|n|$.

The eigenvalue of the BFKL kernel in the next-to-leading approximation
was calculated initially in QCD (see ref.~\cite{FL}).
It contains the contributions proportional to
the
Kronecker symbols $\delta _{n,0}$ and $\delta _{n,2}$. But in $N=4$ SUSY
these nonanalytic terms are  cancelled and a simple expression having the property of
the hermitian separability was obtained~\cite{trajN4, KL}.
Furthermore, the final result in two loops is a sum of special
functions having the property of maximal transcendentality~\cite{KL}.
In a different context, one-loop anomalous
dimension matrix for twist-2 operators in this model was calculated and its
eigenvalues turned out to be proportional to $\psi (1)-\psi (j-1)$,
which is related to the integrability of the evolution
equation for the quasi-partonic operators in
$N=4$ SUSY~\cite{L4}. The integrability in this model
has also been established for other operators and in higher
loops~\cite{MZ, BS}.

The maximal transcendentality principle suggested
in Ref.~\cite{KL} allowed to extract the
universal anomalous dimension up to three loops in $N=4$
SUSY~\cite{KLV, KLOV} from the QCD results~\cite{VMV}.
This principle was also helpful for finding a closed integral equation
for the cusp anomalous
dimension in this model~\cite{ES, BES} satisfying the AdS/CFT
correspondence~\cite{Malda, GKP, W}. In the framework of the
asymptotic Bethe ansatz with wrapping corrections
the maximal transcendentality principle
gave a possibility to calculate the anomalous
dimension up to five loops~\cite{KLRSV} in an agreement with
the BFKL predictions. Moreover, the intercept of the
BFKL Pomeron at a large 't Hooft coupling constant in $N=4$ SUSY was
found in Refs.~\cite{KLOV, Polch}. Next-to-leading corrections to the
BFKL equation can be obtained with the use of the effective action for
the reggeized gluon interactions~\cite{eff, last}.

A simple ansatz for gluon production amplitudes with the maximal helicity
violation in a planar limit for $N=4$ SUSY  was suggested by Bern,
Dixon and Smirnov~\cite{BDS}. This ansatz for the elastic case at large coupling
was confirmed by Alday and Maldacena~\cite{AM1}.
However, later for the multi-particle production amplitude these authors obtained
the result different from the BDS predictions~\cite{AM2}.
It was shown in ref.~\cite{BLS1}, that already in the 6 point case
the BDS ansatz is in a disagreement with the
Steinmann relations~\cite{Steinmann} which are equivalent to the requirement, that
the production amplitude does not have simultaneous
singularities in overlapping channels. The BDS result was not confirmed also by
direct two loop calculations ~\cite{BernDrum}. The reason
for the breakdown of the BDS ansatz
is related to the fact, that the BDS amplitude for the transition
$2\rightarrow 4$ in
the multi-Regge kinematics does not contain the Mandelstam cut
contribution~\cite{Mandel}.
This new term appears
in the
$j_2$-plane of the $t_2$ channel
at the physical kinematic regions, where the invariants in the direct channels
have the following signs $s,s_2>0;\,s_1,s_3<0$ or
$s,s_1,s_2,s_3<0;\,s_{012},s_{123}>0$~\cite{BLS1}.
In LLA the cut contribution for the 6-point
amplitude was calculated in LLA with the use of the BFKL equation~\cite{BLS2}. The
corresponding amplitude in the region $s,s_2>0;\,s_1,s_3<0$
can be written in the factorized form
\beq
M_{2\rightarrow 4}=M^{BDS}_{2\rightarrow 4}\,(1+i\Delta _{2\rightarrow 4})
\label{corLLA}
\eeq
where $M^{BDS}_{2\rightarrow 4}$ is the BDS amplitude~\cite{BDS} and
\beq
\Delta _{2\rightarrow 4}=\frac{a}{2}\, \sum _{n=-\infty}^\infty (-1)^n
\int _{-\infty}^\infty \frac{d\nu }{\nu ^2+\frac{n^2}{4}}\,
\left(\frac{q_3^*k^*_a}{k^*_bq_1^*}\right)^{i\nu -\frac{n}{2}}\,
\left(\frac{q_3k_a}{k_bq_1}\right)^{i\nu +\frac{n}{2}}\,
\left(s_2^ {\omega (\nu , n)}-1\right)\,.
\label{LLA}
\eeq
Here $k_a,k_b$ are transverse components of produced gluon momenta,
$q_1,q_2,q_3$ are the momenta of reggeons in the corresponding
crossing channels and
\beq
\omega (\nu , n)=4a\,
\Re \left(2\psi (1)-\psi (1+i\nu +\frac{n}{2})-\psi (1+i\nu -\frac{n}{2})\right).
\label{eigen}
\eeq
The correction $\Delta$ is M\"{o}bius invariant in the momentum space and can be
written in terms of the four-dimensional anharmonic ratios~\cite{BLS2} in an accordance
with the results of refs.~\cite{DrumSmir}.

It was shown also, that in a general case of the Mandelstam
cut corresponding to a composite state of $n$ reggeized
gluons the Hamiltonian coincides with the local Hamiltonian
for an open integrable Heisenberg spin chain~\cite{Intopen}.

In this paper we reproduce some results of ref.~\cite{BLS2} using
general arguments based only on analyticity and factorization of the
6-point amplitude without any unitarity constraints incorporated
in the BFKL approach.
Also the exponentiation ansatz with an additional phase factor
for the BDS amplitude is investigated in LLA.

\section{Dispersion relation in multi-Regge kinematics}

\bigskip\bigskip
The BDS amplitude~\cite{BDS} for the transition $2\rightarrow 3$ in the multi-Regge
kinematics can be
written in the following form compatible
with the Steinmann relation (see \cite{BLS1})
\beq
\frac{M_{2\rightarrow 3}^{BDS}}{\Gamma (t_1)\Gamma (t_2)}=
(-s_1)^{\omega _{12}}(-s\kappa _{12})^{\omega _2}c_1^{12}+
(-s_2)^{\omega _{21}}(-s\kappa _{12})^{\omega _1}c_2^{12}\,,\,\,\kappa _{12}=
|k_a|^2\,,
\eeq
where $\Gamma (t_i)$ are the reggeized gluon residues,  $k_a$ is the transverse momentum of
the produced particle and
we put the normalization point $\mu ^2$ in the Regge factors equal to unity. The gluon Regge
trajectories are
\beq
\omega _r=\omega (|q_r|^2)= -\frac{\gamma_K}{4}\,\ln \frac{|q_r|^2}{\lambda ^2}\,,\,\,
\gamma _K\approx 4a\,,\,\,a=
\frac{g^2\,N_c}{8\pi ^2}\,,\,\,\omega _{12}=\omega _1-\omega _2\,,
\eeq
where $\gamma _K$ is the cusp anomalous dimension and $\lambda ^2=\mu ^2 \exp (1/\epsilon )$
for $D=4-2\epsilon$ with  $\epsilon \rightarrow -0$.
The real coefficients $c_1^{12},\,c_2^{12}$
are given below~\cite{BLS1}
\beq
c_1^{12}=|\Gamma _{12}|\,\frac{\sin \pi (\omega _1-\omega _a)} {\sin \pi \omega _{12}}\,,
\,\,
c_2^{12}=|\Gamma _{12}|\,\frac{\sin \pi (\omega _2-\omega _a)} {\sin \pi \omega _{21}}\,,
\eeq
where the Reggeon-Reggeon-gluon vertex $\Gamma _{12}$ in the physical region $s,s_1,s_2>0$ is
\beq
\Gamma _{12}(\ln \kappa _{12}-i\pi)=|\Gamma _{12}|\,\exp (i\pi \,\omega _a)\,,\,\,
\omega _a =\frac{\gamma _K}{8}\,
\ln \frac{|k_a|^2\lambda ^2}{|q_1|^2|q_2|^2}\,,
\label{omegaa}
\eeq
\beq
\ln |\Gamma _{12}|=\frac{\gamma _K}{4}\, \left(-\frac{1}{4}\ln ^2\frac{|k_a|^2}{\lambda ^2}
-\frac{1}{4}\ln ^2\frac{|q_1|^2}{|q_2|^2}+
\frac{1}{2}\ln \frac{|q_1|^2|q_2|^2}{\lambda ^4}\ln \frac{|k_a^2|}{\mu ^2}
+\frac{5}{4}\zeta _2\right)\,.
\eeq

It is well known, that one particle production
amplitude with the reggeon exchanges having definite signatures $\tau _1,\tau _2=\pm 1$ in
the crossing
channels
$t_1$ and $t_2$ has the factorized form in all physical regions~\cite{Cambridge}
\beq
\frac{M^{\tau _1\tau _2}_{2\rightarrow 3}}{\Gamma (t_1)\Gamma (t_2)}=
|s_1|^{\omega _{1}}\xi _1\,V^{\tau _1\tau _2}
\,|s_2|^{\omega _2}\xi _2\,,\,\,V^{\tau _1\tau _2}=\frac{\xi _{12}}{\xi _1}\,c_1^{12}+
\frac{\xi _{21}}{\xi _2}\,c_2^{12}\,,
\label{sign23}
\eeq
where
\beq
\xi _1=e^{-i\pi \omega _1}-\tau _1\,,\,\,\xi _2=e^{-i\pi \omega _2}-\tau _2\,,\,\,
\xi _{12}=e^{-i\pi \omega _{12}}+
\tau _1\tau _2\,,\,\,\xi _{21}=e^{-i\pi \omega _{21}}+
\tau _1\tau _2\,.
\eeq

Moreover, for two particles production
in the multi-Regge
kinematics the amplitude
with definite signatures
$\tau _i$ in
three crossing channels can be also presented in the factorized form~\cite{Cambridge}

\beq
\frac{M^{\tau _1\tau _2\tau _3}_{2\rightarrow 4}}{\Gamma (t_1)\Gamma (t_3)}=
|s_1|^{\omega _{1}}\xi _1\,V^{\tau _1\tau _2}
\,|s_2|^{\omega _2}\xi _2\,V^{\tau _2\tau _3}\,|s_3|^{\omega _3}\xi _3\,,
\label{sign24}
\eeq
where $V^{\tau _2\tau _3}$ is obtained from $V^{\tau _1\tau _2}$ (\ref{sign23})
with the corresponding
substitutions
\beq
V^{\tau _2\tau _3}=\frac{\xi _{23}}{\xi _2}\,c_1^{23}+
\frac{\xi _{32}}{\xi _3}\,c_2^{23}\,.
\eeq
For the second produced gluon with the transverse momentum $k_b$ the coefficients
$c^{23}$ and phase $\omega _b$ are
\beq
c_1^{23}=|\Gamma _{23}|\,\frac{\sin \pi (\omega _2-\omega _{b})} {\sin \pi \omega _{23}}\,,
\,\,
c_2^{23}=|\Gamma _{23}|\,\frac{\sin \pi (\omega _3-\omega _{b})} {\sin \pi \omega _{32}}\,,
\eeq
\beq
\omega _{b}= \frac{\gamma _K}{8}\ln \frac{|k_b|^2\lambda ^2}{|q_2|^2|q_3|^2}\,,\,\,
|k_b|^2=\left|\frac{s_2s_3}{s_{123}}\right|\,.
\label{omegab}
\eeq
In an accordance with the Steinmann relations
the Regge hypothesis
leads
to the following expression for the Regge pole contribution
$M_{2\rightarrow 4}^{pole}$~\cite{Cambridge, BLS1}
\[
\frac{M_{2\rightarrow 4}^{pole}}{\Gamma (t_1)\Gamma (t_3)}=(-s_1)^{\omega _{12}}\,
(-s_{012}\kappa _{12})^{\omega _{23}}\,(-s\kappa _{12}\kappa _{23})^{\omega _3}\,
c_1^{12}\,c_1^{23}
\]
\[
+(-s_3)^{\omega _{32}}
(-s_{123}\kappa _{23})^{\omega _{21}}\,(-s\kappa _{12}\kappa _{23})^{\omega _1}\,c_2^{12}\,
c_2^{23}
+(-s\kappa _{12}\kappa _{23})^{\omega _2}\,(-s_1)^{\omega _{12}}\,(-s_3)^{\omega _{32}}\,
c_1^{12}\,c_2^{23}
\]
 \[
 +(-s_2)^{\omega _{21}}
 (-s_{012}\kappa _{12})^{\omega _{13}}\,(-s\kappa _{12}\kappa _{23})^{\omega _3}\,
 \frac{\sin \pi \omega _1}{\sin \pi \omega _2}\,
 \frac{\sin \pi \omega _{23}}{\sin \pi \omega _{13}}\,c_2^{12}\,c_1^{23}
 \]
\beq
 +(-s_2)^{\omega _{23}}
 (-s_{123}\kappa _{23})^{\omega _{31}}\,(-s\kappa _{12}\kappa _{23})^{\omega _1}\,
 \frac{\sin \pi \omega _3}{\sin \pi \omega _2}\,
 \frac{\sin \pi \omega _{21}}{\sin \pi \omega _{31}}\,c_2^{12}\,c_1^{23}\,.
\label{Stau24}
 \eeq
 It is valid in all physical regions different by signs of momenta $p_A,p_B,k_1$ and $k_2$.
Using the identity
\beq
\frac{\sin \pi \omega _1}{\sin \pi \omega _2}\,
\frac{\sin \pi \omega _{23}}{\sin \pi \omega _{13}}\,
\frac{\xi _{13}\xi _2}{\xi _{23}\xi _1} +
\frac{\sin \pi \omega _3}{\sin \pi \omega _2} \,
\frac{\sin \pi \omega _{21}}{\sin \pi \omega _{31}}\,
\frac{\xi _{31}\xi_2}{\xi _{21}\xi _3}
=1\,,
\eeq
one can verify the Regge factorization of the signatured amplitudes
$M^{\tau _1\tau _2\tau _3}_{2\rightarrow 4}$ (\ref{sign24}).
Note, that there is another useful relation
\beq
\frac{\sin \pi \omega _1}{\sin \pi \omega _2}\,
\frac{\sin \pi \omega _{23}}{\sin \pi \omega _{13}} +
\frac{\sin \pi \omega _3}{\sin \pi \omega _2} \,
\frac{\sin \pi \omega _{21}}{\sin \pi \omega _{31}}
=1\,.
\eeq

The two-gluon production amplitude in the multi-Regge kinematics can be written as a sum
of the Regge pole and Mandelstam cut contributions~\cite{BLS1}
\beq
M_{2\rightarrow 4}= M_{2\rightarrow 4}^{pole}+M_{2\rightarrow 4}^{cut}\,,
\label{PolCut}
\eeq
where $M_{2\rightarrow 4}^{cut}$ is non-zero only in two kinematic regions
restricted by the inequalities
$s,s_2>0; \,s_1,s_3<0$ and $s,s_1,s_2,s_3<0; \,s_{012},s_{123}>0$.

The pole term (\ref{Stau24}) in the region $s,s_2>0; \,s_1,s_3<0$ is given below
\[
\frac{M_{2\rightarrow 4}^{pole}}{
|s_1|^{\omega _{1}}
|s_2|^{\omega _{2}}
|s_3|^{\omega _{3}}\,\Gamma (t_1)\Gamma (t_3)}=
e^{-i\pi \omega _3}\,
c_1^{12}\,c_1^{23}+
e^{-i\pi \omega _1}\,
c_2^{12}\,
c_2^{23} +
e^{-i\pi \omega _2}\,
c_1^{12}\,
c_2^{23}
\]
 \beq
 +e^{-i\pi \omega _2}\,\left(
 e^{i\pi \omega _{13}}\,
 \frac{\sin \pi \omega _1}{\sin \pi \omega _2}\,
 \frac{\sin \pi \omega _{23}}{\sin \pi \omega _{13}}
 +
 e^{-i\pi \omega _{13}}\,
 \frac{\sin \pi \omega _3}{\sin \pi \omega _2}\,
 \frac{\sin \pi  \omega _{21}}{\sin \pi \omega _{31}}
 \right)\,c_2^{12}\,c_1^{23}\,.
 \eeq

With the use of the relation
  \[
  e^{i\pi \omega _{13}}\,\frac{\sin \pi \omega _1}{\sin \pi \omega _2}\,
  \frac{\sin \pi \omega _{23}}{\sin \pi \omega _{13}} + e^{-i\pi \omega _{13}}\,
 \frac{\sin \pi \omega _3}{\sin \pi \omega _2}
  \frac{\sin \pi \omega _{21}}{\sin \pi \omega _{31}}
  \]
  \beq
  =\cos \pi \omega _{13}+i\,\frac{
  \sin \pi \omega _1\,\sin \pi \omega _{23} +\sin \pi \omega _3 \,\sin \pi \omega _{21}
}{\sin \pi \omega _2}
  \eeq
this result can be simplified
\[
\frac{M_{2\rightarrow 4}^{pole}}{
|s_1|^{\omega _{1}}
|s_2|^{\omega _{2}}
|s_3|^{\omega _{3}}\,|\Gamma _{12}||\Gamma _{23}|\,\Gamma (t_1)\Gamma (t_3)}=
-e^{-i\pi \omega _2}
\frac{\sin \pi \omega _{2a}}{\sin \pi \omega _{12}}
\left(
e^{-i\pi \omega _{1b}}+2i
\frac{\sin \pi \omega _{1}}{\sin \pi \omega _{2}}\sin\pi \omega _{2b}\right)
\]
\beq
+e^{-i\pi \omega _b}\,
\frac{\sin \pi \omega _{1a}}{\sin \pi \omega _{12}}=\frac{2\,\sin \pi \omega _a\sin \pi \omega _b}{i\sin \pi \omega _{2}}+
e^{-i\pi \omega _2}\,e^{i\pi (\omega _a+\omega _b)}\,.
\eeq
We can present $M^{pole}$ in the region $s,s_2>0; \,s_1,s_3<0$
as a sum of three contributions
\[
\frac{M_{2\rightarrow 4}^{pole}}{|s_1|^{\omega _{1}}|s_2|^{\omega _{2}}|s_3|^{\omega _{3}}
  |\Gamma _{12}||\Gamma _{23}|
  \,\Gamma (t_1)\Gamma (t_3)}
=
\frac{2e^{-i\pi \omega _2}\cos \pi \omega _2\,
\sin \pi \omega _a\sin \pi \omega _b}{i\sin \pi \omega _{2}}
\]
\beq
+ie^{-i\pi \omega _2}
\sin \pi (\omega _a+\omega _b)+e^{-i\pi \omega _2}\cos \pi \omega _{ab}\,,\,\,
\omega _{ab}=\frac{\gamma _K}{4}\,\ln \frac{|k_a||q_3|}{|k_b||q_1|}\,.
\label{pol}
\eeq
Here two first terms have the phase structure of the cut contribution
$M_{2\rightarrow 4}^{cut}$ considered below in (\ref{cut}) and can be included in it, which
gives a possibility to redefine $M_{2\rightarrow 4}^{pole}$ in the form
\beq
\frac{M_{2\rightarrow 4}^{pole}}{|s_1|^{\omega _{1}}|s_2|^{\omega _{2}}|s_3|^{\omega _{3}}
  |\Gamma _{12}||\Gamma _{23}|
  \,\Gamma (t_1)\Gamma (t_3)}=e^{-i\pi \omega _2}\cos \pi \omega _{ab}\,.
  \label{pole2}
\eeq
Indeed, in an accordance with the above discussed
representation for planar amplitudes in the multi-Regge kinematics the cut contribution
can be presented as follows (it corresponds to the last two terms in the
pole contribution (\ref{Stau24})) (cf.~\cite{Intopen})
\beq
M_{2\rightarrow 4}^{cut} \sim \left(1-\Phi ^{\omega _{13}}\right)\,
(-s_{012}\kappa _{12})^{\omega _{13}}\,(-s\kappa _{12}\kappa _{23})^{\omega _3}
(-s_2)^{\omega _{21}}\int _{-i\infty}^{i \infty}
\frac{d\omega _{2'}}{2\pi i}\,\phi (\omega _{2'})\,(-s_2)^{\omega _{2'}}\,.
\label{cut1}
\eeq
Here we introduced the quantity $\Phi$ which coincides with the anharmonic ratio
related to the conformal invariance of the production amplitudes in the
momentum space
\beq
\Phi=\frac{ss_2}{s_{012}s_{123}}\,,\,\,1-\Phi \approx \frac{|k_a+k_b|^2}{s_2}
\eeq
and the partial wave $\phi (\omega _{2})$  is real for real $\omega_2$ and depends on
various invariants
in crossing channels.
The above expression for $M_{2\rightarrow 4}^{cut}$ is non-zero only in two regions,
where
$\Phi =\exp (\mp 2\pi i)$ (really this fact fixes the relative coefficient of two terms
at the first factor in (\ref{cut1}). From this representation
we conclude, that the phase structure
of the cut contribution at $s,s_2>0,\,s_1,s_3<0$ (corresponding to
$\Phi =\exp (- 2\pi i)$) is
\beq
\frac{M_{2\rightarrow 4}^{cut}}{|s_1|^{\omega _{1}}|s_2|^{\omega _{2}}|s_3|^{\omega _{3}}
  |\Gamma _{12}||\Gamma _{23}|
  \,\Gamma (t_1)\Gamma (t_3)}= i\,e^{-i\pi \omega _2}  \,
\int _{-i\infty}^{i \infty}
\frac{d\omega _{2'}}{2\pi i}\,f(\omega _{2'})\,e^{- i\pi \omega _{2'}}\,
|s_2|^{\omega _{2'}}\,.
\label{cut}
\eeq
The redefined partial wave $f (\omega _{2'})$ can contain the pole
$\sim 1/\omega _{2'}$,
which allows one to absorb the terms $\sim i \exp (-i\pi \omega _2)$
from $M_{2\rightarrow 4}^{pole}$ to $M_{2\rightarrow 4}^{cut}$, as it was done
in transition from (\ref{pol}) to (\ref{pole2}).

In a similar way the pole and cut contributions in the region $s,s_1,s_2,s_3>0;\,s_{012},s_{123}>0$
($\Phi =\exp (2\pi i)$) can
be presented in the form
\beq
\frac{M_{2\rightarrow 4}^{pole}}{|s_1|^{\omega _{1}}|s_2|^{\omega _{2}}|s_3|^{\omega _{3}}
  |\Gamma _{12}||\Gamma _{23}|
  \,\Gamma (t_1)\Gamma (t_3)}=\cos \pi \omega _{ab}\,,
\eeq
\beq
\frac{M_{2\rightarrow 4}^{cut}}{|s_1|^{\omega _{1}}|s_2|^{\omega _{2}}|s_3|^{\omega _{3}}
  |\Gamma _{12}||\Gamma _{23}|
  \,\Gamma (t_1)\Gamma (t_3)}= -i\,
\int _{-i\infty}^{i \infty}
\frac{d\omega _{2'}}{2\pi i}\,f(\omega _{2'})\,
|s_2|^{\omega _{2'}}\,.
\eeq

\section{Factorization and analytic properties of $M_{2\rightarrow 4}$}

The BDS amplitude in the multi-Regge kinematics for the physical channel in which
$s,s_2>0,\,s_1,s_3<0$ is given below (see ref.~\cite{BLS1})
\beq
\frac{M_{2\rightarrow 4}^{BDS}}{
|s_1|^{\omega _{1}}
|s_2|^{\omega _{2}}
|s_3|^{\omega _{3}}\,|\Gamma _{12}||\Gamma _{23}|\,\Gamma (t_1)\Gamma (t_3)}=
C\,e^{-i\pi \omega _2}\,e^{i\pi (\omega _a+\omega _b)}=e^{-i\pi \omega _2}\,
e^{i\pi \delta }\,,
\label{BDS1}
\eeq
where
\beq
\delta = \frac{\gamma_K}{4}\,
\ln \frac{|q_1||q_2||k_a||k_b|}{|k_a+k_b|^2|q_2|^2}
\label{BDSphase}
\eeq
and we used the following expression for the phase factor $C$
\beq
C= \exp \left( \frac{\gamma_K}{4}\,
 i\pi \ln \frac{|q_1|^2|q_3|^2}{|k_a+k_b|^2\lambda ^2}\right)\,.
\label{Cphase}
\eeq
Note, that the phase $\delta$ does not contain infrared divergencies and can be written
as follows
\beq
\delta = \frac{\gamma_K}{8}\,
\ln \frac{u_2 u _3}{(1-u_1)^2}\,,
\label{BDSphasePhi}
\eeq
where $u _r$ are anharmonic ratios of invariants in the momentum space
\beq
u_1=\Phi = \frac{s\,s_2}{s_{123}s_{012}}\,,\,\,
u_2 = \frac{s_{3}t_1}{s_{123}t_2}
\,,\,\,u_3= \frac{s_{1}t_3}{s_{012}t_2}\,.
\label{phi012}
\eeq
Correspondingly, in the physical region where $s,s_1,s_2,s_3<0;\,s_{012},s_{123}>0$
the BDS amplitude can be written as follows
\beq
\frac{M_{2\rightarrow 4}^{BDS}}{
|s_1|^{\omega _{1}}
|s_2|^{\omega _{2}}
|s_3|^{\omega _{3}}\,|\Gamma _{12}||\Gamma _{23}|\,\Gamma (t_1)\Gamma (t_3)}=
C\,e^{-i\pi \omega _2}\,e^{i\pi (\omega _a+\omega _b)}=
e^{-i\pi \delta }\,,
\label{BDS1a}
\eeq

According to the hypothesis formulated in refs.~\cite{AM2, Drummond:2007bm} the
correct expression for
$M_{2\rightarrow 4}$ can be obtained from $M_{2\rightarrow 4}^{BDS}$ by multiplying it
by a factor $c$ being a function of these anharmonic relations
\beq
M_{2\rightarrow 4}=c \,M_{2\rightarrow 4}^{BDS}\,.
\label{factoriz}
\eeq
The factorization hypothesis together with the above discussed representation of
$M_{2\rightarrow 4}$ in the form of
a sum of the Regge pole and the Mandelstam cut contributions (\ref{PolCut}) leads to the
following relation for $c$ valid
in the region $s,s_2>0,\,s_{1},s_3<0$
\beq
c\,e^{i\pi \,\delta}=\cos \pi \omega _{ab}+i\int _{-i\infty}^{i \infty}\frac{d\omega}{2\pi i}\,
f(\omega )\,e^{-i\pi \omega}\,(1-u_1)^{-\omega}\,,\,\,1-u_1 \approx
\frac{|k_a+k_b|^2}{s_2}\rightarrow +0\,.
\label{anal1}
\eeq
Here $f (\omega )$ is a real function depending on two invariant variables
\beq\label{a111}
\phi _2=\frac{u_2}{1-u_1}\approx
\frac{|q_1|^2|k_b|^2}{|k_a+k_b|^2|q_2|^2}\,,\,\,\phi _3=\frac{u_3}{1-u_1}\approx
\frac{|q_3|^2|k_a|^2}{|k_a+k_b|^2|q_2|^2}\,.
\eeq
The phases $\delta $ and
$\omega _{ab}$ also can be expressed in terms of these variables
\beq
\delta =\frac{\gamma _K}{8}\,\ln (\phi _3\phi _2)\,,\,\,
\omega _{ab}
=\frac{\gamma _K}{8}\,\ln \frac{\phi _3}{\phi _2}\,.
\eeq

In a similar way for the production amplitude in the region $s,s_1,s_2,s_3<0;\,s_{012},s_{123}>0$ one can
derive the relation
\beq
c\,e^{-i\pi \,\delta}=\cos \pi \omega _{ab}-i
\int _{-i\infty}^{i \infty}\frac{d\omega}{2\pi i}\,
f(\omega )\,(u_1-1)^{-\omega}\,,\,\,u_1>1\,.
\label{anal2}
\eeq

The above representations for $c$ are valid
on the second sheets of the Riemann surface of this function at $u_1\rightarrow 1$. In the
quasi-multi-regge kinematics
$s_1,s_3\gg s_2\sim t_1\sim t_2\sim t_3$ the anharmonic ratio $u_1$ is not close to unity.
The first sheet of the Riemann surface for $c$ corresponds to the production amplitude
$M_{2\rightarrow 4}$ in the kinematic region where $s,s_1,s_2,s_3>0$. In this region
the amplitude is
regular at $u_1= 1$ and has the singularity at $u_1=0$.

To illustrate these analytic properties  let us consider the BDS amplitude in the
region
$s,s_1,s_2,s_3>0$.
It contains the following dependence on $u_1$~\cite{BLS1}
\beq
\ln M_{2\rightarrow 4}^{BDS}=-\frac{\gamma _K}{8}\left(Li _2(1-u_1)+\ln u_1\,
\ln (-\sqrt{u_2\,u_3})+\frac{1}{2}\ln ^2u_1\right)+...\,,
\eeq
where we included also the phase $i\pi (\omega _a+\omega _b)$ and used the identity
\beq
\frac{|q_1|\,|q_3|\,|k_1|\,|k_2|}{-s_2\,|q_2|^2}=-\sqrt{u_2\,u_3}\,.
\eeq

With the use of
the integral representation for the dilogarithm function $Li _2(z)$
\beq
Li_2 (z)= -\int _0^z\frac{dx}{x}\,\ln (1-x)=z\int _1^\infty \frac{dz'}{z'(z'-z)}\,\ln z'
\eeq
we conclude, that the one loop BDS amplitude~\cite{BDS} has  singularities at $u_1=0$
\beq
I_6^{(1)}+F_6^{(1)}=-\frac{\gamma _K}{8}\,\int _{-\infty}^0 \frac{(1-u_1)
\,du '_1}{(1-u '_1)
(u _1-u' _1 )}\,\left(\ln (1-u '_1)
-\ln (-\sqrt{u_2\,u _3}) -\ln (-u '_1)\right)+...\,.
\eeq

In the multi-Regge regime, where  $s,s_2>0;\,
s_1,s_3<0$, the invariant $u _1$ is close to unity
\beq
u_1\approx e^{-2\pi i}\left(1-\frac{|k_1+k_2|^2}{s_2}\right)
\eeq
and the amplitude should be continued to the second sheet of the $u _1$-plane through
the lower edge of the cut at $u_1<0$, which generates
the additional term (cf.~\cite{BLS1})
\beq
\Delta \left(I_6^{(1)}+F_6^{(1)}\right)=
-\pi i\,\frac{\gamma _K}{4}\,\left(\ln (1-u_1)
-\ln \frac{t_1t_3}{s_2\lambda ^2}-\ln u _1\right)\,.
\eeq
Note, that this term has also a singularity at $u _1=1$ and is pure
imaginary in the physical region $u _1<1$.

In the next section we consider the two loop contribution. In this case the second  order
expansion of the
BDS exponent on the first sheet also can be presented in a form of the dispersion integral
which relates its
real and imaginary parts. However, this expression
does not agree with the Steinmann
relations.

\section{Two loop production amplitude $M_{2\rightarrow 4}$}
For the production amplitude in LLA the following expression for $M_{2\rightarrow 4}$ in
the region
$s,s_2>0;\,s_1,s_3<0$ was obtained in two loops with the use of the
$s$-channel unitarity~\cite{BLS2}
\beq
M_{2\rightarrow 4}=c\,M_{2\rightarrow 4}^{BDS}\,,\,\,c=1+\frac{a^2}{4}\,r_2+O(a^3)\,,
\eeq
where
\[
r_2\approx Li_2(1-u _1)\,\ln
\frac{(1-u _1)}{u_2}\ln
\frac{(1-u_1)}{u_3}
\]
\beq
+Li_2(1-u _2)\,\ln
 \frac{(1-u _2)}{u_3}\ln
 \frac{(1-u_2)}{u _1}+
 Li_2(1-u _3)\,\ln  \frac{(1-u _3)}{u _2}\ln  \frac{(1-u_3)}{u _1}
\,.
\label{twoloo}
 \eeq
 Here we introduced the four-dimensional anharmonic ratios (\ref{phi012})
and included additional terms to provide the invariance of
$M_{2\rightarrow 4}$ under
the cyclic permutations.
The added contributions are not essential in the multi-Regge kinematics, although for the exact
two-loop
result they are important. Note, that another physical region $s,s_1,s_2,s_3<0;\,s_{012},s_{123}>0$,
where
 $u _1=\exp (2\pi i)$, is also described correctly by the above expression (\ref{twoloo})
for $r_2$.

 We should take into account also a similar cut
 contribution to the transition amplitude $3\rightarrow 3$. But in fact it is
already contained
 in eq. (\ref{twoloo}) due to the
 relations (cf.~\cite{BLS1, BLS2})
 \beq
 \frac{s_{13}s_{02}}{st'_2}=u_2\rightarrow 1+\frac{|q_1+q_3-q_2|^2}{t'_2}\,,\,\,
 \frac{1-u _2}{u_1 }\rightarrow \frac{|q_1+q_3-q_2|^2|q_2|^2}{|q_3|^2|q_1|^2}\,,\,\,
 \frac{u_1 }{u _3}\rightarrow \frac{|q_3|^2|q_1|^2}{|k_2|^2|k_1|^2}\,.
 \eeq
 Thus, our expression (\ref{twoloo}) in two loops leads to the correct multi-Regge asymptotics in all
 channels. Moreover, the conformal invariance in the momentum representation is
 valid also in higher loops of LLA if we substitute the anharmonic ratios in the
 two-dimensional transverse subspace by the corresponding four dimensional ratios
 $u _{2,3}$ and the power of the logarithm $\ln s_2$
at large $s_2$ by the following expression
 \beq
 -2\pi i \frac{\ln ^ns_2}{n} \rightarrow
 (-1)^{n-1}\int _0^{1-u_1 }\frac{dt}{t}\,\ln^{n-1}t\,\ln (1-t)\,,
 \eeq
 which can be written in terms of the polylogarithm function $Li _{n+1}(z)$.

 Let us expand the BDS amplitude in the region $s,s_2>0;\,s_{012},s_{123}<0$ in the perturbation
 series  to investigate a possibility to correct its bad analytic properties
 with the factor $c$ depending on the anharmonic ratios. It can be presented
 at this kinematics in the form~\cite{BLS1}
 \beq
 M_{2\rightarrow 4}^{BDS}=C\,\Gamma (t_1)
 \,(-s_1)^{\omega _{1}}\,\Gamma (\ln \kappa _{12}-i\pi )\,
 (-s_2)^{\omega _2}\, \Gamma (\ln \kappa _{23}-i\pi ) \,
 (-s_3)^{\omega _{3}}\,\Gamma (t_2)\,,
 \eeq
 which was simplified above (see (\ref{BDS1})).
 Note, that the phase $\delta$ (\ref{BDSphase}) does not contain infrared divergencies and
depends on an anharmonic ratio in the
 two-dimensional momentum space. It can be written also in terms of
 four-dimensional anharmonic ratios (\ref{BDSphasePhi}).

 The first order term of the expansion of the phase in (\ref{BDS1}) over $\delta$ corresponds
 to the Mandelstam cut contribution in one-loop approximation~\cite{BLS1}. The second order
term $-\pi ^2\delta ^2/2$ of the phase factor expansion
\beq
e^{i\pi \delta}=1+i\pi \delta -\pi ^2\frac{\delta ^2}{2}+...
\eeq
contradicts the Steinmann relations and analytic properties for
$M_{2\rightarrow 4}$ if we would not take into account the additional
 logarithmic contribution $\sim \ln s_2$ appearing in the factor $c$. On
the other hand, the LLA result for
$c$ in the two loop approximation after its analytic continuation to the
region $s,s_2>0;\,s_1,s_3<0$ can be written as follows~\cite{BLS2}
\beq
c\approx
1-2\pi i \,\frac{a^2}{4}\, \ln s_2
\,\ln \frac{|k_2|^2|q_1|^2}{|k_1+k_2|^2|q_2| ^2} \,\ln
\frac{|k_1|^2|q_3|^2}{|k_1+k_2|^2|q_2| ^2}+...\,.
\eeq
It does not contain the phase factor $\exp (-\pi i)$ in the argument of $\ln s_2$
due to the pure imaginary asymptotics of the function $Li_2(1-u_1)$
in eq. (\ref{twoloo}) at $u_1\rightarrow \exp (-2\pi i)$. To
obtain the correct
real part for $F_{2\rightarrow 4}$ in an accordance with the phase structure of the cut
contribution (\ref{cut1}) depending on the argument $-s_2$ we should find somewhere
the following real term
\beq
 \Delta c=
 -\frac{a^2\,\pi ^2}{2}
\,\ln \frac{|k_2|^2|q_1|^2}{|k_1+k_2|^2|q_2| ^2} \,\ln
\frac{|k_1|^2|q_3|^2}{|k_1+k_2|^2|q_2| ^2}  \,.
\label{Deltac}
\eeq
It is remarkable, that this correction is contained already at the BDS factor in eq. (\ref{factoriz}). Indeed,
$\Delta c$ can be
written as follows
\[
 \Delta c= -\frac{a^2\,\pi ^2}{2}\,\left(\ln ^2\frac{|k_1||k_2||q_1||q_2|}{|k_1+k_2|^2|q_2|}-
\ln ^2\frac{|k_1||q_3|}{|k_2||q_1|}\right)
\]
\beq
\approx
 -\frac{f_{K}^{2}\,\pi ^2}{32}
\left(\ln ^2\frac{|k_1||k_2||q_1||q_2|}{|k_1+k_2|^2|q_2|^2}-
\ln ^2\frac{|k_1||q_3|}{|k_2||q_1|^2}\right)
=-\pi ^2\frac{\delta ^2}{2}+
\pi ^2\frac{\omega _{ab}^2}{2}\,.
\eeq
The first contribution $-\delta ^2/2$ is the second order term in the expansion of the phase
factor
$\exp (i\delta)$ in (\ref{BDS1}) and the second contribution $\omega ^2_{ab}/2$ is opposite in sign
to the second order
term in the expansion of the factor $\cos \omega _{ab}$ included in the pole contribution
$M_{2\rightarrow 4}^{pole}$ (\ref{pole2}). Note, that the phase factor $\exp (-i\pi \omega _2)$
exists in all
three amplitudes $M_{2\rightarrow 4},\,M_{2\rightarrow 4}^{pole}$ and
$M_{2\rightarrow 4}^{cut}$.

Thus, the two-loop result for the two gluon production amplitude in LLA is in a full agreement with
analyticity requirements and a factorization hypothesis (\ref{factoriz}).
In fact it follows completely from these properties without
any necessity to solve the BFKL equation~\cite{BLS2}. Moreover, the BFKL kernel can be calculated from
the two loop correction. Note, that the analytic properties of the cut contribution
(\ref{cut1}) predict the pure imaginary result also for the next-to-leading term in $r_2$ (not
proportional to $\ln s_2$). Recently~\cite{LP} this prediction was confirmed by an analytic continuation
of the exact expression for two loop production
amplitude obtained in refs.~{\cite{DDS, GSVV}}. It means,
that the representation of the six point amplitude in terms of the $P$-exponents~\cite{Drummond:2007bm}
is in an agreement
with the Mandelstam cut asymptoticsat least in two loops~\cite{BLS2}.

In a similar way in the physical region $s,s,s_1,s_2,s_3<0;\,s_{012},s_{123}>0$ the factor $c$ in two
loops can be presented as follows
\beq
c\approx
1+2\pi i \,\frac{a^2}{4}\, (\ln (-s_2)-i\pi )
\,\ln \frac{|k_2|^2|q_1|^2}{|k_1+k_2|^2|q_2| ^2} \,\ln
\frac{|k_1|^2|q_3|^2}{|k_1+k_2|^2|q_2| ^2}+...\,.
\eeq
Here the real term $\sim a^2 \pi ^2$ contradicts the analytic properties for the Mandelstam cut
contribution in this region and it is cancelled as above with the two loop expansions of the BDS phase
and the
pole contribution
\beq
\Delta c=-\pi ^2\frac{\delta  ^2}{2}+\pi ^2\frac{\omega^2 _{ab}}{2}\,.
\eeq

\section{Exponentiation hypothesis}

As it was argued above, in the region $s,s_2>0;\,s_1,s_3<0$ for the multi-loop amplitude
$M_{2\rightarrow 4}$  in LLA and
beyond it one can use the relations
\beq
M_{2\rightarrow 4}=c\,M_{2\rightarrow 4}^{BDS}=
M_{2\rightarrow 4}^{pole}+M_{2\rightarrow 4}^{cut}\,,
\label{expanal}
\eeq
where $c$ is an invariant function of three anharmonic ratios in
the momentum space. The  BDS amplitude is given by eq. (\ref{BDS1}), $M_{2\rightarrow 4}^{pole}$
is known explicitly (see (\ref{pole2})) and
the analytic properties of $M_{2\rightarrow 4}^{cut}$
are defined by the integral (\ref{cut}).

These relations can be considered as  a set of equations for the real functions
$c$ and $f(\omega _2)$ although they seem
to be incomplete, because for example in two loops we can add to the result the next-to-leading correction
of the form (see ref.~\cite{LP})
\beq
\Delta M_{2\rightarrow 4}=ia^2
\,\Delta ^{NLLA} (\Phi_2,\Phi_3),
\eeq
where $\Phi_2$ and $\Phi_3$ are given by eq.~\ref{a111}.

Generalizing the BDS hypothesis one can assume, that the correct amplitude
$M_{2\rightarrow 4}$ has an exponential form. However, it will be shown below,
that the factor $c$
can not be a pure phase  in the region $s,s_2>0, s_1,s_3<0$
\beq
c\ne e^{i\phi}\,.
\label{nonexp}
\eeq
This conclusion is based on
the fact, that in LLA the complex structure of the
production amplitude (including the phase of the BDS ansatz) is known.

We start with the dispersion representation for the cut contribution to the
production amplitude in LLA (see (\ref{cut}))
\beq
\frac{M_{2\rightarrow 4}^{cut}}{|M_{2\rightarrow 4}^{BDS}|}=ia\,\pi \,e^{-i\pi \omega _2}\,\sum
_{n=0}^{\infty}(\ln (-s_2))^n\,c_n\,a^n\,,
\eeq
where the coefficients $c_n$ due to eq. (\ref{LLA}) are known in the form of integrals over $\nu$ and
a sum over $n$ from powers of the eigenvalue (\ref{eigen}) of the BFKL kernel for the adjoint
representation. For the real part
one  obtains with a leading accuracy
\beq
\Re \frac{M_{2\rightarrow 4}^{cut}}{e^{-i\pi \omega _2}|M_{2\rightarrow 4}^{BDS}|}=
a\,\pi ^2\,\sum _{n=0}^{\infty}(\ln (s_2))^{n-1}\,n\,c_n\,a^n\,.
\eeq
We devided the equality with the factor $\exp (-i\pi \omega _2)$
because it is common for all contributions.

On the other hand, using the exponentiation hypothesis with the additional assumption, that
the remainder  function is a phase
\beq
c=e^{i\phi }\,,\,\,\phi \approx \Delta_{2\rightarrow 4}\,,
\label{exphypot}
\eeq
where $\Delta _{2\rightarrow 4}$ is given in eq. (\ref{LLA}),
one can obtain the coefficients
$c_n$ for $n\ge 2$ from the expansion
\beq
\Re \frac{M_{2\rightarrow 4}^{cut}}{e^{-i\pi \omega _2}|M_{2\rightarrow 4}^{BDS}|}=-\frac{\pi
^2}{2}\,
\left(a\sum _{k=0}^\infty (\ln s_2)^k\,c_k\,a^k \right)^2\,,
\eeq
where
\beq
c_0 =\ln \frac{|q_1|^2|q_3|^2}{|k_1+k_2|^2\lambda ^2}
+\frac{1}{2}\,\ln \frac{|k_1|^2\lambda ^2}{|q_1|^2|q_2|^2}
+\frac{1}{2}\,\ln \frac{|k_2|^2\lambda ^2}{|q_3|^2|q_2|^2}
=
\frac{1}{2}\,\ln \frac{|q_1|^2|q_3|^2|k_1|^2|k_2|^2}{|k_1+k_2|^4|q_2| ^4}\,.
\eeq
Here the first term appears from the factor $C$ (\ref{Cphase}) and two last terms
are from the phases $\omega _a$ (\ref{omegaa}) and $\omega _b$ (\ref{omegab}).
For the coefficient $c_1$ we have from the previous section (see (\ref{Deltac}))
\beq
c_1=-\frac{1}{2} \ln \frac{|k_2|^2|q_1|^2}{|k_1+k_2|^2|q_2| ^2} \,\ln
\frac{|k_1|^2|q_3|^2}{|k_1+k_2|^2|q_2| ^2}\,.
\eeq
Thus, from the exponentiation hypothesis (\ref{exphypot}) we obtain the recurrent relation for
$c_n$ at $n\ge 2$
\beq
n\,c_n=-\frac{1}{2}\,\sum _{k=0}^{n-1}c_k\,c_{n-1-k}\,.
\eeq
In particular,
\beq
c_2=\frac{1}{8}\,\ln \frac{|q_1|^2|q_3|^2|k_1|^2|k_2|^2}{|k_1+k_2|^4|q_2| ^4}\,
\ln \frac{|k_2|^2|q_1|^2}{|k_1+k_2|^2|q_2| ^2} \,\ln
\frac{|k_1|^2|q_3|^2}{|k_1+k_2|^2|q_2| ^2}\,.
\label{c2rec}
\eeq
Let us introduce the generating function $y(x)$
\beq
\frac{M_{2\rightarrow 4}^{cut}}{e^{-i\pi \omega _2}|M_{2\rightarrow 4}^{BDS}|}=ia\,\pi
\,y(a\ln
(-s_2))\,,\,\,
y(x)=\sum _{n=0}^{\infty}x^n\,c_n\,.
\eeq
This function satisfies the equation
\beq
\frac{d}{d\,x}\,y(x)=-\frac{1}{2}\,y^2(x)+b\,,
\eeq
where
\beq
y(0)=c_0\,,\,\,b=c_1+
\frac{c_0^2}{2}=\frac{1}{2}\,\ln ^2\frac{|k_2q_1|}{|k_1q_2|}\,.
\eeq
Its solution is
\beq
y=\sqrt{2b}\,\tanh \left(\sqrt{\frac{b}{2}}x+\delta \right)\,\,,\,\,\,
\coth \delta =\frac{c_0}{\sqrt{2b}}
\eeq
with the perturbative expansion
\beq
y= c_0+c_1\,x-\frac{c_0\,c_1}{2}\,x^2+...\,.
\eeq

This result based on analytic properties of the production amplitude and
on the assumption of the
exponentiation (\ref{exphypot}) of $i\Delta _{2\rightarrow 4}$ in
(\ref{corLLA})  is in an disagreement
with the perturbative solution
(\ref{LLA}) of the BFKL equation corresponding to the factorization
property of
$t$-channel partial waves. In
particular, the
exponent $\exp (-\sqrt{2b}\,x)$ depends only on the module of the
anharmonic ratio $u_2/u_3$ whereas
the correct BFKL expression depends also on a phase.
It turns out, that already in three loops the leading logarithmic result
(\ref{LLA}) contains the special
functions $Li_3(x)$ and $Li_2(x)$ absent in $c_3$. Indeed, according to
ref.~\cite{LP2} the function
$\widetilde{c_2}$ obtained from eq. (\ref{LLA}) has the form
\[
\widetilde{c_2}=\frac{1}{8}\left(2\ln |w|^2\ln ^2|1+w|^2-\frac{4}{3}\ln
^3|1+w|^3
-\frac{1}{2}\ln^2|w|^2\ln |1+w|^2\right)
\]
\beq
+\frac{1}{8}\ln  |w|^2\left(Li_2(-w)+Li_2(-w^*)\right)-\frac{1}{4}
\left(Li_3(-w)-Li_3(-w^*)\right)\,,\,\,w=\frac{q_3k_1}{k_2q_1}\,.
\eeq

Thus, the factor $c$ in the region $s,s_2;\,s_1,s_3<0$ can not be a pure phase in the
physical regions with $u_1=\exp (\pm 2\pi i)$, where the amplitude contains the Mandelstam cuts.
The analytic properties of $c$ are presented in eqs. (\ref{anal1}) and (\ref{anal2}). They show,
that the knowledge of the amplitude in LLA allows one to
calculate not only leading corrections to the imaginary part of the factor $c$, but also -
leading corrections to its real part suppressed by the extra factor $\sim a$
(see ref.~\cite{LP2}).

\section{Conclusion}
In this paper we investigated analytic properties of the planar six point amplitude for $N=4$ SUSY
in the multi-Regge kinematics. This amplitude has the Regge pole and the Mandelstam cut
contributions and should satisfy the Steinmann relations. We  calculated the two loop correction
to the amplitude $M_{2\rightarrow 4}$ at
the region $s,s_2>0;\,s_1,s_3<0$ in an agreement with the results of the paper~\cite{BLS2}
using only analyticity constraints and a factorization hypothesis.
It was shown, that in the next-to-leading approximation the two loop correction to the
factor $c$ in front of the BDS expression should be also pure imaginary. This prediction is confirmed
by direct calculations in ref.~{\cite{LP}}. We also demonstrated above, that
in upper loops the factor $c$ in the Regge kinematics can not be a pure phase
(see (\ref{nonexp})), because such phase structure
would contradict the $t$-channel Regge factorization incorporated in the
BFKL equation. Really the amplitude in LLA can be reproduced completely from
the BDS ansatz with the use of the analyticity and the Regge factorization. 

I thank J. Bartels and A. Prygarin for helpful discussions and the Institute of Advance Studies
of the Tel Aviv University for the kind invitation to stay there in October - December of 2009
when a significant part of this investigation was done. This work was supported by the Russian
grant RFBR-10-02-01338-a.

\end{document}